\documentstyle[11pt,newpasp,twoside,epsf]{article}
\newcommand{\jcomp}{\rmfamily{J. Comp. Phys.}}

\newcommand{\beq}{\begin{equation}}
\newcommand{\eeq}{\end{equation}}
\newcommand{\bdm}{\begin{displaymath}}
\newcommand{\edm}{\end{displaymath}}

\def\edcomment#1{\iffalse\marginpar{\raggedright\sl#1\/}\else\relax\fi} 
\reversemarginpar 
 
\begin{document}
\title{Strong Shocks and Supersonic Winds in Inhomogeneous
Media} 

\author{A.Y. Poludnenko\altaffilmark{1}, A. Frank\altaffilmark{2},
E.G. Blackman
\altaffilmark{3}}
\affil{Department of Physics and Astronomy,\\
       University of Rochester, Rochester, NY 14627-0171}
\altaffiltext{1}{wma@pas.rochester.edu}
\altaffiltext{2}{afrank@pas.rochester.edu}
\altaffiltext{3}{blackman@pas.rochester.edu}
 
\begin{abstract} 
Many astrophysical flows occur in inhomogeneous media. The broad-line
regions (BLR) of active galactic nuclei (AGNs) are one of the
important examples where emission-line clouds interact with the
outflow.

We present results of a numerical study of the interaction of a
steady, planar shock / supersonic postshock flow with a system of
embedded cylindrical clouds in a two-dimensional geometry. Detailed
analysis shows that the interaction of embedded inhomogeneities with
the shock / postshock wind depends primarily on the thickness of the
cloud layer and the arrangement of the clouds in the layer, as opposed
to the total cloud mass and the total number of individual clouds.
This allows us to define two classes of cloud distributions: thin and
thick layers. We define the critical cloud separation along the
direction of the flow and perpendicular to it. This definition allows
us to distinguish between the interacting and noninteracting regimes
of cloud evolution. Finally we discuss mass-loading in such systems.
\end{abstract}

\section{INTRODUCTION}

Mass flows are important in many astrophysical systems from stars to
the most distant active galaxies. Virtually all mass flow studies
focus on homogeneous media. However, the typical astrophysical medium
is inhomogeneous with the inhomogeneities (clumps) arising due to
initial fluctuations of mass distribution, the action of
instabilities, variations in the flow source, etc. The presence of
inhomogeneities can introduce not only quantitative but also
qualitative changes to the overall dynamics of the flow.

Active galactic nuclei represent one astrophysical site where a shock
/ wind interaction with a system of inhomogeneities may take place.
Practically all current models of AGNs agree that the emission-line
clouds in BLRs are essential for explaining the observed properties of
AGNs (Urry \& Padovani 1995; Elvis 2000). However, despite the fact
that the nature of BELR clouds is an open question, any
self-consistent model of the AGNs should properly account for
properties of BELR cloud interaction with mass outflow. In particular,
such self-consistency is important in the context of cloud survival
time and cloud displacement prior to destruction. Even when clouds are
magnetically confined and stabilized against evaporation (Rees 1987),
disruptive action of outflows may prevent cloud survival over the
dynamically significant AGN timescales.

(Klein, McKee, Colella 1994) (hereafter KMC) addressed the problem of
shock-cloud interaction, providing a detailed description of the
dynamics of a single, dense, unmagnetized cloud interacting with a
strong, steady, planar shock. That work gives an excellent
introduction to the subject, in particular the description of the
astrophysical significance of the problem of a shock wave interacting
with a dense cloud (see also (Gregori et al. 2000)). In this work we
investigate the general properties of strong shock / supersonic wind
interaction with a system of embedded clouds and determine the key
quantities governing the evolution of such systems.

\section{RESULTS}

\begin{figure}
\plotfiddle{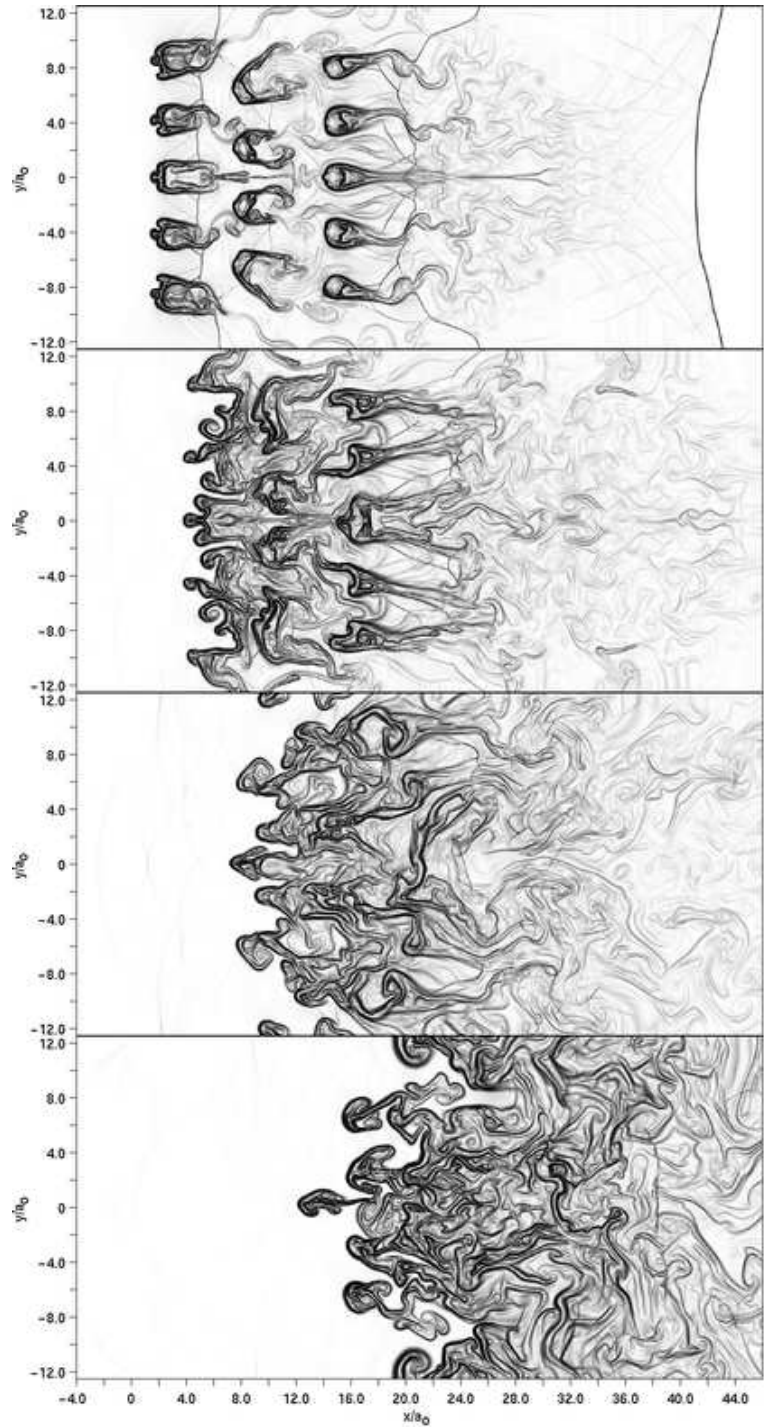}{525 pt}{0}{68}{68}{-148}{0}
\caption{Run $M14$. Time evolution of a system, containing fourteen 
identical clouds in a regular distribution and interacting with a
$M_{S}=10$ shock wave. Shown are synthetic Schlieren images of the
system at times $22 \ t_{SC}$, $35 \ t_{SC}$, $50 \ t_{SC}$, $69 \
t_{SC}$.
\label{14clump}}
\end{figure}

We have numerically investigated the interaction of a strong, planar
shock wave with a system of dense inhomogeneities, embedded in a more
tenuous and cold ambient medium. Our code is the AMRCLAW package,
which implements an adaptive mesh refinement algorithm for the
equations of gas dynamics (Berger \& Oliger 1984; Berger \& Jameson
1985; Berger \& Colella 1989; Berger \& LeVeque 1998) in two
dimensions. We have assumed constant conditions in the global
postshock flow constraining the maximum size of the clouds only by the
condition of the shock front planarity. Our results are applicable to
strong global shocks with Mach numbers $3 \la M_{S} \la 1000$.  The
range of the applicable cloud - unshocked ambient medium density
contrast values is $10 - 1000$. Figure 1 shows a case of a strong
shock and a supersonic post-shock flow interacting with an
inhomogeneous system of fourteen identical clouds in regular
distribution.

Cloud evolution due to the interaction with the global shock and the
postshock flow has four major phases, namely \emph{the initial
compression phase, the re-expansion phase, the destruction phase,} and
finally \emph{the mixing phase}. Each image in Figure 1 roughly
illustrates each of those phases.

The timescale we use to define time intervals in our numerical
experiments is the time it takes for the incident shock wave to sweep
across an individual cloud. This is called the \emph{shock-crossing
time,} $t_{SC}=(2a_{max}/v_{S})$, where $a_{max}$ is the maximum cloud
radius in the distribution.

We define the \emph{cloud destruction time} $t_{CD}$ as the time
when the largest cloud fragment contains less than 50$\%$ of the
initial cloud mass. Typically, in our simulations $t_{CD} \approx
24t_{SC}$.

A simple model for the cloud acceleration during the first three
phases, i.e. prior to its destruction, can be developed. We find the
cloud velocity
\beq
v_{C}(t)= \left\{ \begin{array}{ll}
\displaystyle v_{PS}\Bigg(1-\Big(\frac{t}{t_{SC}}
a_{1}+ a_{2}\Big)^{-1}\Bigg), & t \le 12t_{SC} \\
\displaystyle v_{PS}\Bigg(1-\Big(\bigg(\frac{t}{t_{SC}}-12\bigg)^2b_{1}
+\frac{t}{t_{SC}}a_{1} + a_{2}\Big)^{-1}\Bigg), & t \le t_{CD}
\end{array} \right.
\label{vcnum}
\eeq
Here, $v_{PS}$ is the unperturbed postshock flow velocity. For the
cases of infinitely strong shocks, i.e. $M_{S} \to \infty$, the above
coefficients have the values
\beq
a_{1} \ = \ 1.83 \cdot 10^{-3}; \ a_{2} \ = \ 1.09; \ 
b_{1} \ = \ 8.51 \cdot 10^{-5}.
\label{a1a2Minf}
\eeq
Equation (1) shows that the maximum cloud velocity is not more than
$10\%$ of the global shock velocity and not more than $13\%$ of the
postshock flow velocity.

Cloud displacement prior to its destruction can be described as
\beq
L_{C}(t)= \left\{ \begin{array}{ll}
\displaystyle a_{0}c_{1}\Bigg(\frac{t}{t_{SC}}-
\frac{1}{a_{1}} \ ln\Big(\frac{t}{t_{SC}}\Big(\frac{a_{1}}{a_{2}}
\Big)+1\Big)\Bigg),& 0 \le t \le 12t_{SC} \\
\displaystyle a_{0}c_{1}\Bigg(\frac{t}{t_{SC}}-
c_{2} \ tan^{-1}\Big(\frac{\frac{t}{t_{SC}}-12}{\frac{t}{t_{SC}}c_{4}
+ c_{5}}\Big)-c_{3}\Bigg), & 12t_{SC} \le t \le t_{CD}
\end{array} \right.
\label{lcdnum}
\eeq
In the limiting case $M_{S}\rightarrow \infty$ the values of the
coefficients $a_{1}$ and $a_{2}$ are defined in (2) and $c_{1} \ = \
1.5; \ c_{2} \ = \ 103.22; \ c_{3} \ = 10.9; \ c_{4} \ = \ 9.43\cdot
10^{-2}; \ c_{5} \ = \ 112.56.$ 

We find the maximum cloud displacement prior to its destruction,
$L_{CD}$, is
\beq
L_{CD}=L_{C}(t_{CD}) \le 3.5a_{max}.
\label{lcd}
\eeq
The results of our model are in excellent agreement with the numerical
experiments. The difference in the values of cloud velocity and
displacement between analytical and numerical results is $\la 10\%$.

The principal conclusion of the present work is that the set $\Lambda$
of all possible cloud distributions can be subdivided into two large
subsets $\Lambda_{I}$, \emph{thin-layer systems}, and $\Lambda_{M}$,
\emph{thick-layer systems}, defined as
\beq
\begin{array}{c}
\Lambda_{I} : (\Delta x_{N})_{max} \le L_{CD}, \\
\Lambda_{M} : (\Delta x_{N})_{max} > L_{CD},
\end{array}
\label{2classes}
\eeq
where $(\Delta x_{N})_{max}$ is the maximum cloud separation in the
system along the direction of the flow, or the cloud layer
thickness.

Distributions from each subset exhibit striking similarity in
behaviour (e.g. Figure 2). The systems containing from one to five
clouds, arranged in a single layer, exhibit exactly the same rate of
momentum transfer from the global flow to the clouds. The two fourteen
cloud runs $M_{14}$ and $M_{14r}$ have different cloud distributions,
different total cloud mass, different cloud sizes. Nevertheless, the
rate of the kinetic energy fraction increase during compression and
re-expansion is different from the single layer cases but is still the
same for both fourteen cloud runs. Other global properties exhibit the
same behaviour. We conclude that the evolution of a system of shocked
clouds depends primarily on the total thickness of the cloud layer and
the cloud distribution in it, as opposed to the total number of clouds
or the total cloud mass present in the system.

The key parameters determining the type of evolution, are \emph{the
cloud destruction length} $L_{CD}$, defined above in (3), and
\emph{the critical cloud separation} transverse to the flow
$d_{crit}$. The latter is defined by the condition that the time for
adjacent clouds to expand laterally and merge into a single coherent
structure is equal to the cloud destruction time $t_{CD}$. It can be
described by the expression
\beq
d_{crit}=2a_{0}\Bigg\{\frac{t_{CD}-t_{CC}}{t_{SC}}
\Bigg(\frac{F_{c1}F_{st}}{\chi}\Bigg)^{\frac{1}{2}}\Bigg(\frac{3\gamma 
(\gamma - 1)}{\gamma + 1}\Bigg)^{\frac{1}{2}}+1\Bigg\}.
\label{dcrit}
\eeq

We should be able to determine, either from observations or from
theoretical analysis, the thickness of the cloud layer $(\Delta
x_{N})_{max}$. This determines the class of the given cloud
distribution, $\Lambda_{I}$ or $\Lambda_{M}$. For distributions from
the set $\Lambda_{I}$ with average cloud separation $\langle
\Delta y_{N} \rangle > d_{crit}$ evolution of the clouds during the
compression, re-expansion, and destruction phases will proceed in the
\emph{noninteracting regime} and the formalism for a single cloud
interaction with a shock wave (e.g. KMC) can be used to describe the
system. On the other hand, if the cloud separation is less than
critical, the clouds in the layer will merge into a single structure
before their destruction is completed, and though the compression
phase still can be considered independently for each cloud, evolution
during the re-expansion and destruction phases proceeds in the
\emph{interacting regime}.

\begin{figure}
\plotone{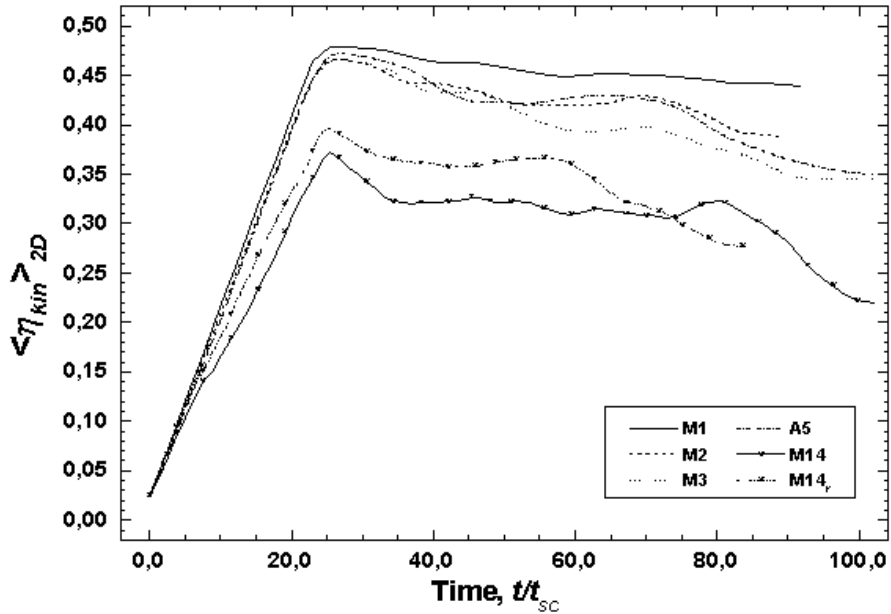}
\caption{Time evolution of the global average of the kinetic energy fraction
$\langle\eta_{kin}\rangle_{2D}$ for the following runs: \\ $M1$, $M2$,
$M3$, $A5$ - systems with 1, 2, 3, 5 identical clouds correspondingly,
arranged in a single row with constant cloud center separation of
$4.0a_{0}$; $M14$ - system with 14 identical clouds with cloud center
separation in a row equal to $4.0a_{0}$, separation between rows equal
to $7.0a_{0}$; $M14_{r}$ - system with 14 clouds of random size and
placed in random positions.
\label{ekin}}
\end{figure}

When the distribution belongs to the subset $\Lambda_{M}$ it is
necessary to determine the average cloud separation projected onto the
direction of the flow $\langle \Delta x_{N} \rangle$ and compare it
against $L_{CD}$: if $\langle \Delta x_{N} \rangle > L_{CD}$ evolution
of the system can be roughly approximated as of a set of distributions
from the subset $\Lambda_{I}$ and the above ``thin-layer case''
analysis applies. If, on the other hand, $\langle \Delta x_{N} \rangle
\le L_{CD}$ (especially if $\langle \Delta y_{N} \rangle < d_{crit}$)
the system evolution is dominated by cloud interactions and a thin
layer formalism is inapplicable.

Finally we consider mass-loading. Here our principal conclusion is
that mass-loading is not significant in the cases of strong shocks and
supersonic winds interacting with inhomogeneities whose density
contrast is in the range $10-1000$. In part this is due to short
survival times of clouds as well as the very low mass loss rates of
the clouds even during the times prior to their destruction.
Therefore, mass-loading in such systems is not likely to have any
appreciable effect on the overall dynamics of the global flow.

The major limitation of our current work is the purely hydrodynamic
nature of our analysis that does not include any consideration of the
magnetic fields. As was discussed by KMC, cold dense inhomogeneities
(clouds) embedded in more tenuous hotter medium are inherently
unstable against the dissipative action of diffusion and thermal
conduction. Although weak magnetic fields, that are dynamically
insignificant up to the moment of cloud destruction, can inhibit
thermal conduction and diffusion and, therefore, stabilize the system
of clouds, those magnetic fields may become dynamically important due
to turbulent amplification during the mixing phase (see (Gregori et
al. 2000) for a three-dimensional study of the wind interaction with a
single magnetized cloud). We intend to provide a fully
magnetohydrodynamic description of the interaction of a strong shock
with a system of clouds in future work.

\acknowledgements

This work was supported in part by the NSF AST-0978765, NASA NAG5-8428,
and DOE DE-FG02-00ER54600 grants.
\\
\\
The most recent results and animations of the numerical experiments,
described above and not mentioned in the current paper, can be found
at \\ www.pas.rochester.edu$/^{\sim}$wma.

\end{document}